\documentclass[conference]{IEEEtran}

\usepackage[english]{babel}
\usepackage{cite}
\usepackage{graphicx}
\graphicspath{{Figures/}}
\usepackage{amsfonts,amsmath,amssymb}
\usepackage{cases}
\usepackage{mathtools}
\usepackage[caption=false,font=footnotesize]{subfig}
\usepackage{url}
\usepackage{siunitx}
\usepackage{newtxtext,newtxmath}
\usepackage{booktabs}

\def\BibTeX{{\rm B\kern-.05em{\sc i\kern-.025em b}\kern-.08em
		T\kern-.1667em\lower.7ex\hbox{E}\kern-.125emX}}

\DeclareMathOperator{\sign}{sign}
\DeclareMathOperator{\modfun}{mod}
\DeclareMathOperator{\wrap}{w}
\DeclareMathOperator{\bigO}{\mathcal{O}}

\DeclarePairedDelimiter\abs{\lvert}{\rvert}
\DeclarePairedDelimiter\norm{\lVert}{\rVert}
\DeclarePairedDelimiterX{\inner}[2]{\langle}{\rangle}{#1, #2}

\newcommand{\eps}{\varepsilon}

\newcommand{\CJ}{\mathcal{J}}
\newcommand{\CR}{\mathcal{R}}
\newcommand{\CC}{\mathcal{C}}
\newcommand{\CA}{\mathcal{A}}
\newcommand{\CS}{\mathcal{S}}
\newcommand{\CM}{\mathcal{M}}

\newcommand{\tpwm}{\varepsilon}
\newcommand{\umax}{u_{m}}

\title{\LARGE \bf Sensorless rotor position estimation by PWM-induced signal injection}
\author{Dilshad Surroop\textsuperscript{1,2}, Pascal Combes\textsuperscript{2}, Philippe Martin\textsuperscript{1} and Pierre Rouchon\textsuperscript{1}
	\thanks{\textsuperscript{1}~D.~Surroop, P.~Martin and P.~Rouchon are with the Centre Automatique et Systèmes, MINES ParisTech, PSL Research University, Paris, France
		{\tt\footnotesize\{dilshad.surroop, philippe.martin,pierre.rouchon\}@mines-paristech.fr}}%
	\thanks{\textsuperscript{2}~D.~Surroop and P.~Combes  are with Schneider Toshiba Inverter Europe, Pacy-sur-Eure, France
		{\tt\footnotesize  pascal.combes@se.com}}
}

\begin{document}
\maketitle

\begin{abstract}
	We demonstrate how the rotor position of a PWM-controlled PMSM can be recovered from the measured currents, by suitably using the excitation provided by
	the PWM itself. This provides the benefits of signal injection, in particular the ability to operate even at low velocity, without the drawbacks of an external probing signal. We illustrate the relevance of the approach by simulations and experimental results.
\end{abstract}


\section*{Nomenclature}
\begin{IEEEdescription}[\IEEEsetlabelwidth{$L_d, L_q$}]
	\item[PWM] Pulse Width Modulation	
	\item[$x^{dq}$] Vector $(x^d, x^q)^T$ in the $dq$ frame
	\item[$x^{\alpha\beta}$] Vector $(x^\alpha, x^\beta)^T$ in the $\alpha\beta$ frame
	\item[$x^{abc}$] Vector $(x^a, x^b, x^c)^T$ in the $abc$ frame
	\item[$R_s$] Stator resistance
	\item[$\mathcal{J}$] Rotation matrix with angle $\pi/2$: 
	$\begin{psmallmatrix} 0 & -1 \\ 1 & 0\end{psmallmatrix}$
	\item[$J$] Moment of inertia
	\item[$n$] Number of pole pairs
	\item[$\omega$] Rotor speed
	\item[$T_l$] Load torque
	\item[$\theta$, $\widehat\theta$] Actual, estimated rotor position
	\item[$\phi_m$] Permanent magnet flux
	\item[$L_d, L_q$] d and q-axis inductances
	\item[$\CC$] Clarke transformation: 
	$\frac{2}{3}\begin{psmallmatrix}1 & -1/2 & -1/2\\ 0 & \sqrt{3}/2 & -\sqrt{3}/2\end{psmallmatrix}$
	\item[$\CR(\theta)$] Rotation matrix with angle $\theta$: $\begin{psmallmatrix}
	\cos \theta & -\sin \theta \\
	\sin \theta & \cos \theta
	\end{psmallmatrix}$
	\item[$\tpwm$] PWM period
	\item[$u_m$] PWM amplitude
	\item[$S(\theta)$] Saliency matrix
	\item[$\bigO$] ``Big O'' symbol of analysis: 
	$k(z,\varepsilon) =\bigO(\varepsilon)$ means $\norm{k(z,\varepsilon)}\leq C\varepsilon$, for some
	$C$ independent  of $z$ and $\varepsilon$.
\end{IEEEdescription}

\section{Introduction}

Sensorless control of AC motors in the low-speed range is a challenging task. Indeed, the observability of the system from the measurements of the currents degenerates at standstill, which limits the performance at low speed of any fundamental-model-based control law.

One now widespread method to overcome this issue is the so-called signal injection technique. It consists in superimposing a fast-varying signal to the control law. This injection creates ripple on the current measurements which carries information on the rotor position if properly decoded. Nonetheless, introducing a fast-varying signal increases acoustic noise and may excite mechanical resonances. For systems controlled through Pulse Width Modulation (PWM), the injection frequency is moreover inherently limited by the modulation frequency. That said, inverter-friendly waveforms can also be injected to produce the same effect, as in the so-called INFORM method~\cite{Schro1996IAS,RobeischlS2004}.  For PWM-fed Permanent Magnet Synchroneous Motors (PMSM), the oscillatory nature of the input may be seen as a kind of generalised rectangular injection on the three input voltages, which provides the benefits of signal injection, in particular the ability to operate even at low velocity, without the drawbacks of an external probing signal.

We build on the quantitative analysis developed in~\cite{SurroopCMR2020arXiv} to demonstrate how the rotor position of a PWM-controlled PMSM can be recovered from the measured currents, by suitably using the excitation provided by the PWM itself. No modification of the PWM stage nor injection a high-frequency signal as~in\cite{WangX2004TPE} is required.

The paper runs as follows: we describe in section~\ref{sec:model} the effect of PWM on the current measurements along the lines of~\cite{SurroopCMR2020arXiv}, slightly generalizing to the multiple-input multiple output framework.  In section~\ref{sec:strategies}, we show how the rotor position can be recovered for two PWM schemes schemes, namely standard single-carrier PWM and interleaved PWM. The relevance of the approach is illustrated in section~\ref{sec:results} with numerical and experimental results.

\section{Virtual measurement induced by PWM}
\label{sec:model}
Consider the state-space model of a PMSM in the $dq$ frame
\begin{subequations}
	\label{eq:system}
\begin{IEEEeqnarray}{rCl}
	\frac{d\phi_s^{dq}}{dt} &=& u_s^{dq} - R_s \imath_s^{dq} - \omega \CJ \phi_s^{dq}, \\
	\frac{J}{n}\frac{d\omega}{dt} &=& n \imath_s^{dq^T}\CJ \phi_s^{dq} - T_l, \\
	\frac{d\theta}{dt} &=& \omega,
\end{IEEEeqnarray}
\end{subequations}
where $\phi_s^{dq}$ is the stator flux linkage, $\omega$ the rotor speed, $\theta$ the rotor position,  $\imath_s^{dq}$ the stator current, $u_s^{dq}$ the stator voltage, and $T_l$ the load torque; $R_s$, $J$, and $n$ are constant parameters (see nomenclature for notations). For simplicity we assume no magnetic saturation, i.e. linear current-flux relations
\begin{subequations}
	\label{eq:currentFlux}
	\begin{IEEEeqnarray}{rCl}
	L_d\imath_s^d &=& \phi_s^d - \phi_m\\
	L_q\imath_s^q &=& \phi_s^q,
	\end{IEEEeqnarray}
\end{subequations}
with $\phi_m$ the permanent magnet flux; see~\cite{JebaiMMR2016IJC} for a detailed discussion of magnetic saturation in the context of signal injection.
The input is the voltage $u_s^{abc}$ through the relation
\begin{IEEEeqnarray}{rCl}\label{eq:input}
u_s^{dq} = \CR(-\theta) \CC u_s^{abc}.
\end{IEEEeqnarray}
In an industrial drive, the voltage actually impressed is not directly~$u_s^{abc}$, but its PWM encoding $\CM\bigl(u_s^{abc},\frac{t}{\tpwm}\bigr)$, with $\tpwm$ the PWM period. The function~$\CM$ describing the PWM is 1-periodic and mean~$u_s^{abc}$ in the second argument, i.e. $\CM\bigl(u_s^{abc},\tau+1\bigr)=\CM\bigl(u_s^{abc},\tau\bigr)$ and $\int_0^1\CM\bigl(u_s^{abc},\tau\bigr)\, d\tau = u_s^{abc}$; its expression is given in section~\ref{sec:strategies}. Setting  $s_0^{abc}\bigl(u_s^{abc},\sigma\bigr):=\mathcal{M}(u_s^{abc},\sigma)-u_s^{abc}$, the impressed voltage thus reads
\begin{IEEEeqnarray*}{rCl}
u_\text{pwm}^{abc} &=& u_s^{abc}+s_0^{abc}\Bigl(u_s^{abc},\frac{t}{\tpwm}\Bigr),
\end{IEEEeqnarray*}
where $s_0^{abc}$ is 1-periodic and zero mean in the second argument; $s_0^{abc}$ can be seen as a PWM-induced rectangular probing signal, which creates ripple but has otherwise no effect. Finally, as we are concerned with sensorless control, the only measurement is the current $\imath_s^{abc} = \CC^T \CR(\theta)\imath_s^{dq}$, or equivalently $\imath_s^{\alpha\beta} = \CR(\theta) \imath^{dq}_s$ since $\imath^a_s + \imath^b_s + \imath^c_s = 0$.

A precise quantitative analysis of signal injection is developed in~\cite{CombeJMMR2016ACC,SurroopCMR2020arXiv}. Slightly generalizing these results to the multiple-input multiple-output case, the effect of PWM-induced signal injection can be analyzed thanks to second-order averaging in the following way.
Consider the system 
\begin{IEEEeqnarray*}{rCl}
	\dot x &=& f(x) + g(x)\bigl(u + s_0(u,\tfrac{t}{\varepsilon})\bigr), \\
	y &=& h(x),
\end{IEEEeqnarray*}
where $u$ is the control input, $\varepsilon$ is a the (assumed small) PWM period, and $s_0$ is 1-periodic in the second argument, with zero mean in the second argument; then we can extract from the actual measurement~$y$ with an accuracy of order $\varepsilon$ the so-called \emph{virtual measurement} (see \cite{CombeJMMR2016ACC,SurroopCMR2020arXiv})
\begin{IEEEeqnarray*}{rCl}
	y_v(t) &:=& h'\bigl(x(t)\bigr) g\bigl(x(t)\bigr) \CA\bigl(u(t)\bigr),
\end{IEEEeqnarray*}
i.e. we can compute by a suitable filtering process an estimate
\begin{IEEEeqnarray*}{rCl}
\widehat{y_v}(t) &=& y_v(t)+\bigO(\eps).
\end{IEEEeqnarray*}
The matrix $\CA$, which can be computed online, is defined by
\begin{IEEEeqnarray*}{rCl}
	\CA(\upsilon) &:=& \int_0^1 s_1(\upsilon,\tau)s_1^T(\upsilon,\tau)\, d\tau,
\end{IEEEeqnarray*}
where $s_1$ is the zero-mean primitive in the second argument of $s_0$, i.e.
\begin{IEEEeqnarray*}{rCl}
s_1(\upsilon,\tau) &:=& \int_0^1 s_0(\upsilon,\sigma)\,d\sigma - \int_0^1\int_0^\tau s_0(\upsilon,\sigma)\,d\sigma d\tau.
\end{IEEEeqnarray*}
The quantity $\eps h'\bigl(x(t)\bigr) g\bigl(x(t)\bigr)s_1\bigl(u(t),\tfrac{t}{\varepsilon}\bigr)$ is the ripple caused on the output~$y$ by the excitation signal
$s_0\bigl(u(t),\tfrac{t}{\varepsilon}\bigr)$; though small, it contains valuable information when properly processed.

For the PMSM~\eqref{eq:system}--\eqref{eq:input} with output~$\imath_s^{\alpha\beta}$, some algebra yields
\begin{IEEEeqnarray*}{rCl}
y_v &=& \begin{bmatrix}
\CR(\theta)\begin{psmallmatrix}\frac{1}{L_d}& 0\\ 0& \frac{1}{L_q}\end{psmallmatrix} & 0_{2\times1} & \CR'(\theta)\imath_{dq}
\end{bmatrix}\!
\begin{bmatrix}\CR(-\theta)\CC\\ 0_{1\times2}\\ 0_{1\times2}\end{bmatrix}\CA^{abc}(u^{abc})\\
&=& \CS(\theta) \CC \CA^{abc}(u^{abc}),
\end{IEEEeqnarray*}
where
\begin{IEEEeqnarray*}{rCl}
	\CA^{abc}(\upsilon^{abc}) &:=& \int_0^1 s_1^{abc}(\upsilon^{abc},\tau)s_1^{abc^T}(\upsilon^{abc},\tau)\, d\tau,
\end{IEEEeqnarray*}
and $\CS(\theta)$ is the so-called saliency matrix introduced in~\cite{JebaiMMR2016IJC},
\begin{IEEEeqnarray*}{rCl}
		\CS(\theta) = \frac{L_d + L_q}{2L_d L_q} \begin{pmatrix}
		1 + \frac{L_q-L_d}{L_d + L_q} \cos 2\theta & \frac{L_q-L_d}{L_d + L_q} \sin 2\theta \\
		\frac{L_q-L_d}{L_d + L_q} \sin 2\theta  & 1 - \frac{L_q-L_d}{L_d + L_q} \cos 2\theta
	\end{pmatrix}.
\end{IEEEeqnarray*}
If the motor has sufficient geometric saliency, i.e. if $L_d$ and $L_q$ are sufficiently different, the rotor position~$\theta$ can be extracted  from~$y_v$ as explained in section~\ref{sec:strategies}. When geometric saliency is small, information on $\theta$ is usually still present when magnetic saturation is taken into account, see~\cite{JebaiMMR2016IJC}.

\section{Extracting $\theta$ from the virtual measurement}
\label{sec:strategies}

Extracting the rotor position~$\theta$ from~$y_v$ depends on the rank of the $2\times3$ matrix $\CC\CA^{abc}(\upsilon^{abc})$. The structure of this matrix, hence its rank, depends on the specifics of the PWM employed. After recalling the basics of single-phase PWM, we study two cases: standard three-phase PWM with a single carrier, and three-phase PWM with interleaved carriers.

Before that, we notice that $\CC\CA^{abc}(\upsilon^{abc})$ has the same rank as the $2\times2$ matrix
\begin{IEEEeqnarray*}{rClCl}
\CA^{\alpha\beta}(\upsilon^{abc}) &:=& \CC\CA^{abc}(\upsilon^{abc})\CC^T\\
&=& \int_0^1 s_1^{\alpha\beta}(\upsilon^{abc},\tau)s_1^{\alpha\beta^T}(\upsilon^{abc},\tau)\, d\tau,
\end{IEEEeqnarray*}
where $s_1^{\alpha\beta}(\upsilon^{abc},\tau) := \CC s_1^{abc}(\upsilon^{abc},\tau)$. Indeed,
\begin{IEEEeqnarray*}{rCl}
\CA^{\alpha\beta}(\upsilon^{abc})\CA^{\alpha\beta^T}(\upsilon^{abc}) &=& 
\CC\CA^{abc}(\upsilon^{abc})\CC^T\CC\CA^{abc^T}(\upsilon^{abc})\CC^T\\
&=& \CC\CA^{abc}(\upsilon^{abc})\bigl(\CC\CA^{abc}(\upsilon^{abc})\bigr)^T,
\end{IEEEeqnarray*}
which means that $\CA^{\alpha\beta}(\upsilon^{abc}$ and $\CC\CA^{abc}(\upsilon^{abc})$ have the same singular values, hence the same rank.
There is thus no loss of information when considering $\CS(\theta)\CA^{\alpha\beta}(u^{abc})$ instead of the original virtual measurement~$y_v$.

\subsection{Single-phase PWM}
\begin{figure}
	\includegraphics{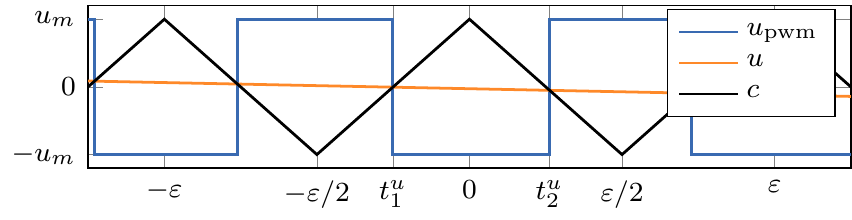}
	\caption{PWM: $u$ is compared to $c$ to produce $u_{\rm pwm}$}
	\label{fig:pwm}
\end{figure}

In ``natural'' PWM with period~$\varepsilon$ and range~$[-\umax,\umax]$, the input signal~$u$ is compared to the $\varepsilon$-periodic triangular carrier
\begin{IEEEeqnarray*}{rCl}
c(t) &:=& 
\begin{dcases*} 
\umax+4\wrap\bigl(\tfrac{t}{\varepsilon}\bigr) & if $-\frac{\umax}{2}\le\wrap\bigl(\frac{t}{\varepsilon}\bigr)\le0$\\ 
\umax-4\wrap\bigl(\tfrac{t}{\varepsilon}\bigr) & if $0\le\wrap(\frac{t}{\varepsilon})\le\frac{\umax}{2}$;
\end{dcases*}
\end{IEEEeqnarray*}
the 1-periodic function $\wrap(\sigma):=\umax\modfun(\sigma+\frac{1}{2},1)-\frac{\umax}{2}$ wraps the normalized time~$\sigma=\frac{t}{\varepsilon}$ to $[-\frac{\umax}{2},\frac{\umax}{2}]$. If $u$ varies slowly enough, it crosses the carrier~$c$ exactly once on each rising and falling ramp, at times $t_1^u<t_2^u$ such that
\begin{IEEEeqnarray*}{rCl}
u(t_1^u) &=& \umax+4\wrap\Bigl(\frac{t_1^u}{\varepsilon}\Bigr)\\
u(t_2^u) &=& \umax-4\wrap\Bigl(\frac{t_2^u}{\varepsilon}\Bigr).
\end{IEEEeqnarray*}
\begin{figure}
	\centering
	\hspace*{0.6em}\includegraphics{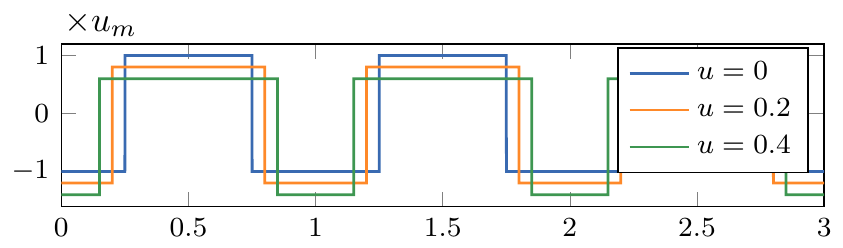}
	\includegraphics{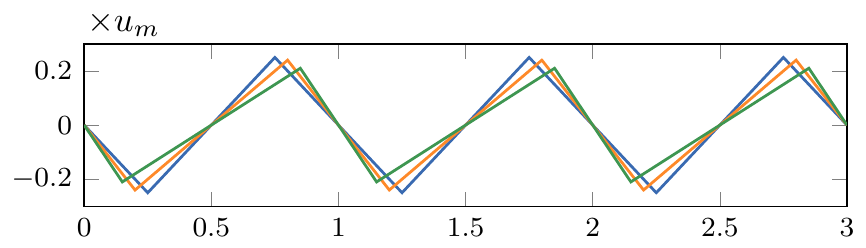}
	\includegraphics{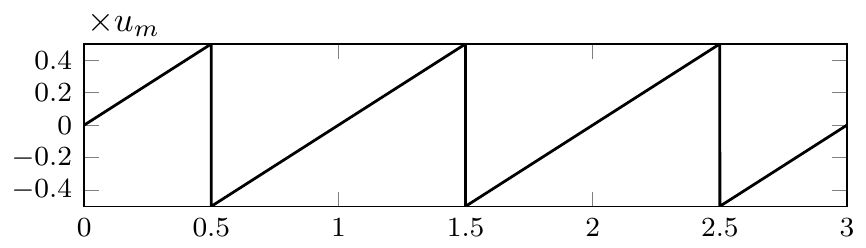}
	\caption{$s_0(u,\cdot)$ (top) and $s_1(u,\cdot)$ (middle) for $u=0, 0.2, 0.4$; $\wrap$ (bottom).}
	\label{fig:s0s1w}
\end{figure}
The PWM-encoded signal is therefore given by
\begin{IEEEeqnarray*}{rCl}
u_{\rm pwm}(t)&=&
\begin{dcases*} 
\umax & if $-\frac{\umax}{2}<\wrap\bigl(\frac{t}{\varepsilon}\bigr)\le\wrap\bigl(\frac{t_1^u}{\varepsilon}\bigr)$\\ 
-\umax & if $\wrap\bigl(\frac{t_1^u}{\varepsilon}\bigr)<\wrap\bigl(\frac{t}{\varepsilon}\bigr)\le\wrap\bigl(\frac{t_2^u}{\varepsilon}\bigr)$\\ 
\umax & if $\wrap\bigl(\frac{t_2^u}{\varepsilon}\bigr)<\wrap\bigl(\tfrac{t}{\varepsilon}\bigr)\le\frac{\umax}{2}$.
\end{dcases*}
\end{IEEEeqnarray*}
Fig.~\ref{fig:pwm} illustrates the signals $u$, $c$ and~$u_{\rm pwm}$.
The function 
\begin{IEEEeqnarray*}{rCl}
\mathcal{M}(u,\sigma)&:=& 
\begin{dcases*} 
\umax & if $-2\umax<4\wrap(\sigma)\le u-\umax$\\ 
-\umax & if $u - \umax<4\wrap(\sigma)\le \umax -u$\\ 
\umax & if $\umax-u<4\wrap(\sigma)\le2\umax$
\end{dcases*}\\
&=& \umax+\umax\sign\bigl(u-\umax-4\wrap(\sigma)\bigr)\\
&&\quad+\>\umax\sign\bigl(u-\umax+4\wrap(\sigma)\bigr),
\end{IEEEeqnarray*}
which is obviously 1-periodic and with mean~$u$ with respect to its second argument, therefore completely describes the PWM process since 
$u_{\rm pwm}(t)=\mathcal{M}\bigl(u(t),\frac{t}{\varepsilon}\bigr)$.

The induced zero-mean probing signal is then
\begin{IEEEeqnarray*}{rCl}
s_0(u,\sigma) &:=& \mathcal{M}(u,\sigma)-u\\
&=& \umax-u+\umax\sign\bigl(\tfrac{u-\umax}{4} - \wrap(\sigma)\bigr)\\
&&\quad+\>\umax\sign\bigl(\tfrac{u-\umax}{4}+\wrap(\sigma)\bigr),
\end{IEEEeqnarray*}
and its zero-mean primitive in the second argument is
\begin{IEEEeqnarray*}{rCl}
s_1(u,\sigma) &:=& \bigl(1-\tfrac{u}{\umax}\bigr)\wrap(\sigma) - \abs[big]{\tfrac{u-\umax}{4}-\wrap(\sigma)}+\abs{\tfrac{u-\umax}{4}+\wrap(\sigma)}.
\end{IEEEeqnarray*}
The signals $s_0$, $s_1$ and~$w$ are displayed in Fig.~\ref{fig:s0s1w}. Notice that by construction $s_0(\pm\umax,\sigma)=s_1(\pm\umax,\sigma)=0$, so there is no ripple, hence no usable information, at the PWM limits.


\subsection{Three-phase PWM with single carrier}\label{subsec:th_non_shifted}
In three-phase PWM with single carrier, each component $u_s^k$, $k=a,b,c$, of $u_s^{abc}$ is compared to the same carrier, yielding
\begin{IEEEeqnarray*}{rCl}
s_0^k(u_s^{abc},\sigma) &:=& s_0(u_s^k,\sigma)\\
s_1^k(u_s^{abc},\sigma) &:=& s_1(u_s^k,\sigma),
\end{IEEEeqnarray*}
with $s_0$ and $s_1$ as in single-phase PWM. This is the most common PWM in industrial drives as it is easy to implement. 

\begin{figure}[t]
	\centering
	\includegraphics{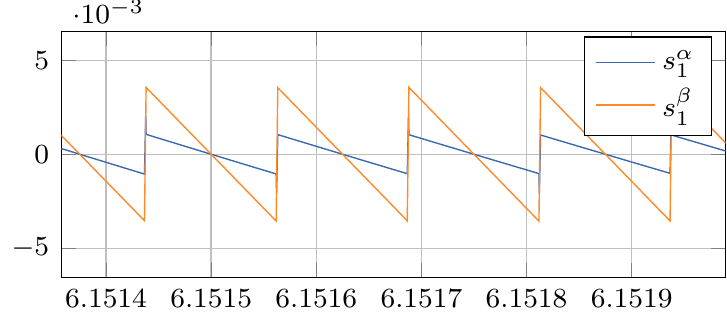}
	\includegraphics{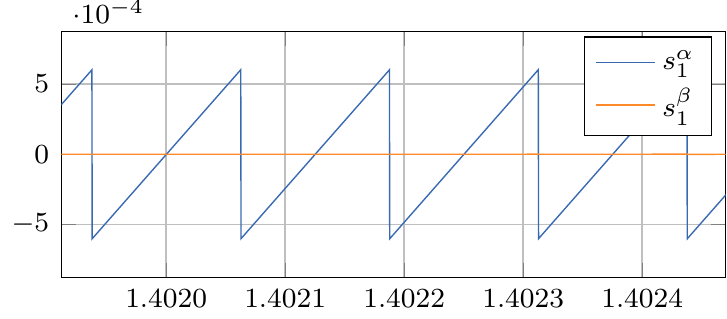}
	\caption{$s_1^{\alpha\beta}$ for single-carrier PWM (simulation data): nondegenerate (top), degenerate (bottom).}
	\label{fig:s1ABsingleCarrier}
\end{figure}
Notice that if exactly two components of~$u_s^{abc}$ are equal, for instance $u_s^c=u_s^b\neq u_s^a$, then 
\begin{IEEEeqnarray*}{rClCl}
s_1^c(u_s^{abc},\sigma) &=& s_1^b(u_s^{abc},\sigma) &\neq& s_1^a(u_s^{abc},\sigma),
\end{IEEEeqnarray*}
which implies in turn that $\CA^{\alpha\beta}(u^{abc})$ has rank~1 (its determinant vanishes); it can be shown this is the only situation that results in rank~1.
If all three components of~$u_s^{abc}$ are equal, then $\CA^{\alpha\beta}(u^{abc})$ has rank~0 (i.e. all its entries are zero); this is a rather exceptional condition that we rule out here. Otherwise $\CA^{\alpha\beta}(u^{abc})$ has rank~2 (i.e. is invertible).
Fig.~\ref{fig:s1ABsingleCarrier} displays examples of the shape of~$s_1^{\alpha\beta}$, in the rank~2 case (top), and in the rank~1 case where~$u_s^c=u_s^b\neq u_s^a$.

As the rank~1 situation very often occurs, it must be handled by the procedure for extracting~$\theta$ from $\CS(\theta)\CA^{\alpha\beta}(u^{abc})$. This can be done by linear least squares, thanks to the particular structure of $\CS(\theta)$. Setting
\begin{IEEEeqnarray*}{rCl}
\begin{pmatrix}\lambda & \mu\\ \mu & \nu\end{pmatrix} &:=& \CA^{\alpha\beta}(u^{abc})\\
\begin{pmatrix}y_{11} & y_{12}\\ y_{21} & y_{22} \end{pmatrix} &:=& \frac{2L_dL_q}{L_d+L_q} y_v.
\end{IEEEeqnarray*}
and $L:=\frac{L_d+L_q}{L_q-L_d}$, we can rewrite  $y_v = \CS(\theta)\CA^{\alpha\beta}(u^{abc})$ as
\begin{equation*}
\underbrace{
	\begin{pmatrix}
	\lambda  & \mu \\
	\mu & \nu \\
	-\mu & \lambda \\
	-\nu & \mu 
	\end{pmatrix}
}_{:= P}
\begin{pmatrix}\cos2\theta \\ \sin2\theta\end{pmatrix} = 
 L\underbrace{
 	\begin{pmatrix}y_{11} - \lambda\\ y_{12} - \mu\\ y_{21} - \mu\\ y_{22} - \nu\\ \end{pmatrix}.
}_{:= d}
\end{equation*} 
The least-square solution of this (consistent) overdetermined linear system is
\begin{IEEEeqnarray*}{rCl}
\begin{pmatrix}\cos2\theta \\ \sin2\theta\end{pmatrix} 
&=& L\bigl[P^TP\bigr]^{-1} P^T d\\
&=& \frac{L}{\lambda^2 + 2\mu^2 + \nu^2}P^T d\\
&=& \tfrac{L}{\lambda^2 + 2\mu^2 + \nu^2}
\begin{pmatrix}\lambda y_{11} + \mu (y_{12} - y_{21}) - \nu y_{22} - \lambda^2 + \nu^2\\ 
\mu(y_{11}+y_{22})+\nu y_{12}+\lambda y_{21}-2\mu(\lambda+\nu)\end{pmatrix}.
\end{IEEEeqnarray*}

Estimates $\widehat{\cos2\theta},\widehat{\sin2\theta}$ for $\cos2\theta,\sin2\theta$ are obtained with the same formulas,
using instead of the actual $y_{ij}$ the estimated
\begin{IEEEeqnarray*}{rClCl}
\begin{pmatrix}\widehat{y_{11}} & \widehat{y_{12}}\\ \widehat{y_{21}} & \widehat{y_{22}} \end{pmatrix} 
&:=& \frac{2L_dL_q}{L_d+L_q} \widehat{y_v}
&=& \frac{2L_dL_q}{L_d+L_q}y_v+\bigO(\eps).
\end{IEEEeqnarray*}
We thus have
\begin{IEEEeqnarray*}{rCl}
\widehat{\cos2\theta} &:=& 
L\frac{\lambda\widehat{y_{11}}+\mu(\widehat{y_{12}}-\widehat{y_{21}})-\nu\widehat{y_{22}}-\lambda^2+\nu^2}{\lambda^2+2\mu^2+\nu^2}\\
&=& \cos2\theta + \bigO(\varepsilon)\\
\widehat{\sin2\theta} &:=& 
L\frac{\mu(\widehat{y_{11}}+\widehat{y_{22}})+\nu \widehat{y_{12}}+\lambda \widehat{y_{21}}-2\mu(\lambda+\nu)}{\lambda^2+2\mu^2+\nu^2}\\
&=& \sin2\theta + \bigO(\varepsilon).
\end{IEEEeqnarray*}
Finally, we get an estimate~$\widehat{\theta}$ of~$\theta$ by
\begin{IEEEeqnarray*}{rClCl}
\widehat\theta &:=& \frac{1}{2}\operatorname{atan2}(\widehat{\sin2\theta},\widehat{\cos2\theta}) + k\pi
&=& \theta + \bigO(\varepsilon), 
\end{IEEEeqnarray*}
where $k \in \mathbb{N}$ is the number of turns.

\subsection{Three-phase PWM with interleaved carriers}\label{subsec:shifted}
\begin{figure}
	\centering
	\includegraphics[width=\columnwidth]{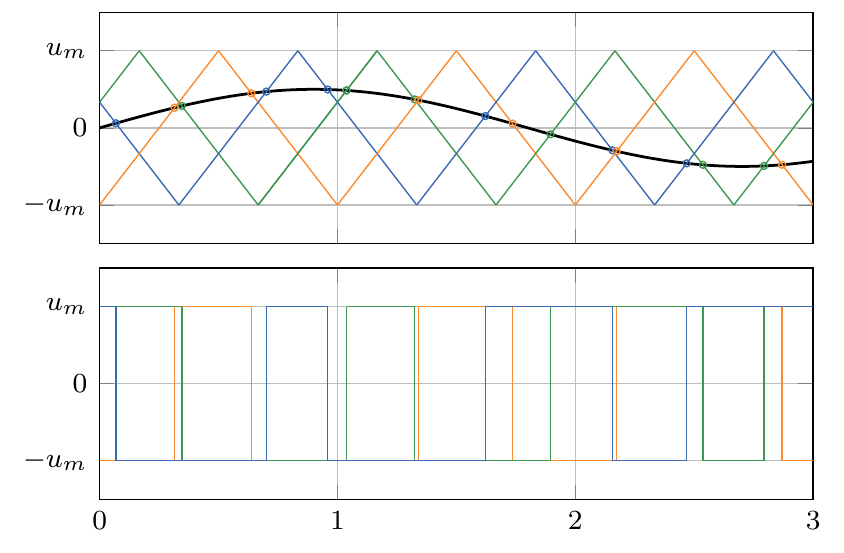}
	\caption{Interleaved carriers (green, orange, blue). The same reference (black) produces different PWM signals.}
	\label{fig:shifted_carriers}
\end{figure}

\begin{figure}
	\centering
	\includegraphics{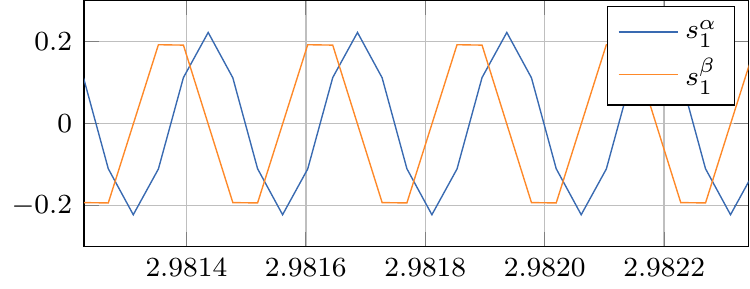}
	\caption{$s_1^{\alpha\beta}$ for interleaved PWM (simulation data).}
	\label{fig:s1ABinterleaved}
\end{figure}

At the cost of a more complicated implementation, it turns out that a PWM scheme with (regularly) interleaved carries offers several benefits over single-carrier PWM. In this scheme, each component of $u_s^{abc}$ is compared to a shifted version of the same triangular carrier (with shift 0 for axis~$a$, $1/3$ for axis~$b$, and $2/3$ for axis~$c$), yielding
\begin{IEEEeqnarray*}{rClrCl}
s_0^a(u_s^{abc},\sigma) &:=& s_0\bigl(u_s^a,\sigma\bigr)&
s_1^a(u_s^{abc},\sigma) &:=& s_1\bigl(u_s^a,\sigma\bigr)\\
s_0^b(u_s^{abc},\sigma) &:=& s_0\bigl(u_s^b,\sigma-\tfrac{1}{3}\bigr)&\qquad
s_1^b(u_s^{abc},\sigma) &:=& s_1\bigl(u_s^b,\sigma-\tfrac{1}{3}\bigr)\\
s_0^c(u_s^{abc},\sigma) &:=& s_0\bigl(u_s^c,\sigma-\tfrac{2}{3}\bigr)&
s_1^c(u_s^{abc},\sigma) &:=& s_1\bigl(u_s^c,\sigma-\tfrac{2}{3}\bigr).
\end{IEEEeqnarray*}
Fig.~\ref{fig:shifted_carriers} illustrates the principle of this scheme. 
Fig.~\ref{fig:s1ABinterleaved} displays an example of the shape of~$s_1^{\alpha\beta}$, which always more or less looks like two signals in quadrature.

Now, even when two, or even three, components of~$u_s^{abc}$ are equal, $\CA^{\alpha\beta}(u^{abc})$ remains invertible (except of course at the PWM limits), since each component has, because of the interleaving, a different PWM pattern. It is therefore possible to recover all four entries of the saliency matrix~$\CS(\theta)$ by
\begin{IEEEeqnarray*}{rClCl}
	\widehat{\CS(\theta)} &:=& \widehat{y_v}\cdot\bigl[\CA^{\alpha\beta}(u^{abc})\bigr]^{-1} &=& \CS(\theta)  + \bigO(\varepsilon). 
\end{IEEEeqnarray*}
Notice now that thanks to the structure of~$\CS(\theta)=(s_{ij})_{ij}$, the rotor angle~$\theta$ can be computed from the matrix entries by
\begin{IEEEeqnarray*}{rCl}
s_{12}+s_{21} &=& \frac{L_q-L_d} {L_dL_q}\sin2\theta\\
s_{11}-s_{22} &=& \frac{L_q-L_d} {L_dL_q}\cos2\theta\\
\theta &=& \frac{1}{2}\operatorname{atan2}(s_{12}+s_{21},s_{11}-s_{22}) + k\pi,
\end{IEEEeqnarray*}
where $k \in \mathbb{N}$ is the number of turns. An estimate~$\widehat{\theta}$ of~$\theta$ can therefore be computed from the entries $(\widehat{s_{ij}})_{ij}$ of~$\widehat{\CS(\theta)}$ by
\begin{IEEEeqnarray*}{rClCl}
	\widehat\theta &=& \frac{1}{2}\operatorname{atan2}(\widehat{s_{12}}+\widehat{s_{21}},\widehat{s_{11}}-\widehat{s_{22}}) + k\pi
&=& \theta + \bigO(\varepsilon),
\end{IEEEeqnarray*}
without requiring the knowledge the magnetic parameters $L_d$ and~$L_q$, which is indeed a nice practical feature.

\section{Simulations and experimental results}
\label{sec:results}
The demodulation procedure is tested both in simulation and experimentally. All the tests, numerical and experimental, use the rather salient PMSM with parameters listed in Table~\ref{table:parameters_PMSM}. The PWM frequency is~\SI{4}{\kilo\hertz}.

The test scenario is the following: starting from rest at t=\SI{0}{\second}, the motor remains there for \SI{0.5}{\second}, then follows a velocity ramp from 0 to \SI{5}{\hertz}~(electrical), and finally stays at \SI{5}{\hertz} from $t=\SI{8.5}{\second}$; during all the experiment, it undergoes a constant load torque of about \SI{40}{\percent} of the rated torque. As this paper is only concerned with the estimation of the rotor angle~$\theta$, the control law driving the motor is allowed to use the measured angle. Besides, we are not yet able to process the data in real-time, hence the data are recorded and processed offline.

\begin{table}[ht]
	\caption{Rated parameters}
	\centering
	\begin{tabular}{lr}
		\toprule
		Rated power & 400 W \\
		Rated voltage (RMS) & 400 V \\
		Rated current (RMS)& 1.66 A \\
		Rated speed & 1800 RPM \\
		Rated torque & 2.12 N.m \\
		\midrule
		Number of pole pairs $n$ & 2 \\
		Stator resistance $R_s$ & 4.25 $\Omega$\\
		$d$-axis inductance $L_d$ &43.25 mH\\
		$q$-axis inductance $L_q$ &69.05 mH \\
		\bottomrule
	\end{tabular}
	\label{table:parameters_PMSM}
\end{table}

\subsection{Single carrier PWM.}
\begin{figure}
	\centering
	\includegraphics{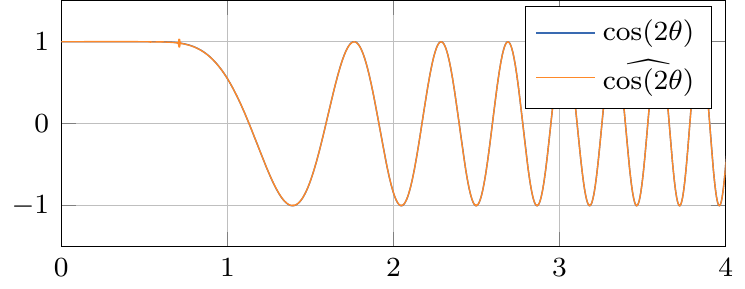}
	\includegraphics{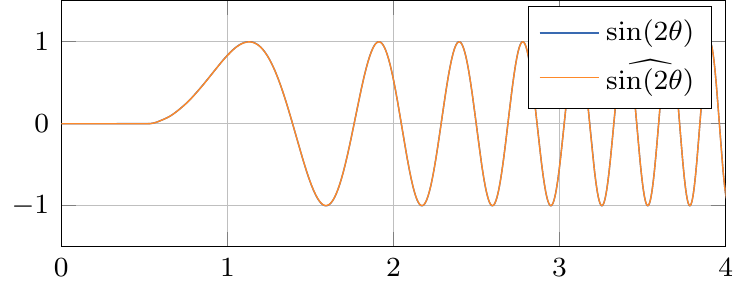}
	\caption{Reconstruction of $\cos2\theta$ and $\sin2\theta$ (simulation).}
	\label{fig:std_cos_sin}
\end{figure}
\begin{figure}
	\centering
	\includegraphics{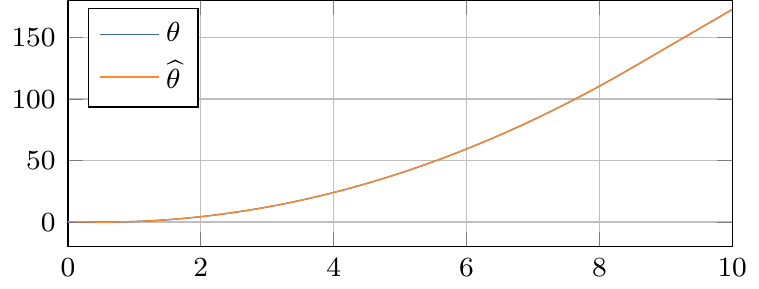}
	\hspace*{0.4cm}\includegraphics{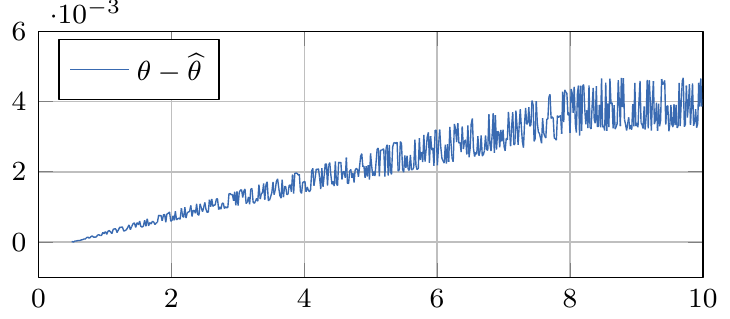}
	\caption{Reconstruction of $\theta$ in \si{\radian} (simulation).}
	\label{fig:std_pos}
\end{figure}

The results obtained in simulation by the reconstruction procedure of section~\ref{subsec:th_non_shifted} for $\cos2\theta$, $\sin2\theta$, and $\theta$, are shown in Fig.~\ref{fig:std_cos_sin} and Fig.~\ref{fig:std_pos}. The agreement between the estimates and the actual values is excellent.

\begin{figure}
	\centering
	\hspace*{0.48cm}\includegraphics{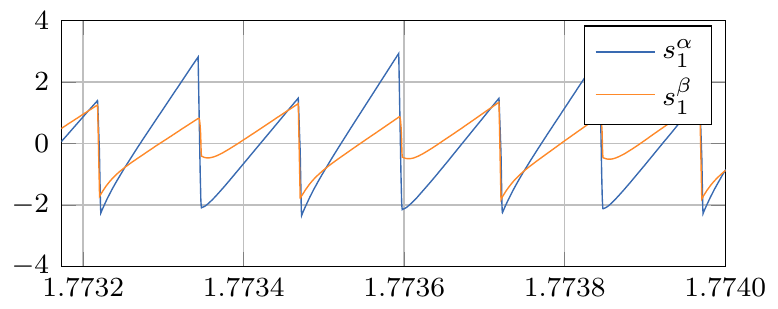}
	\includegraphics{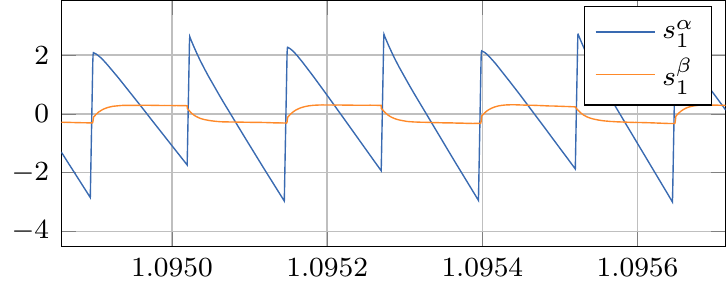}
	\caption{$s_1^{\alpha\beta}$ for single-carrier PWM (experimental data): nondegenerate (top), degenerate (bottom).}
	\label{fig:xp_zoom_s1albe}
\end{figure}

\begin{figure}
	\centering
	\includegraphics{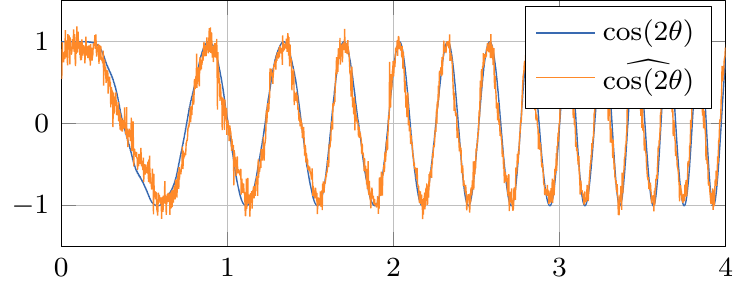}
	\includegraphics{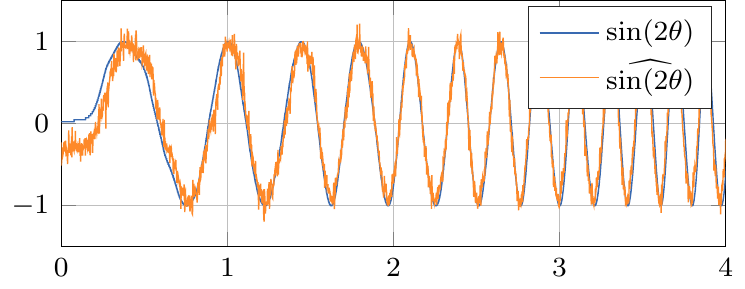}
	\caption{Reconstruction of $\cos2\theta$ and $\sin2\theta$ (experiment).}
	\label{fig:xp_cos_sin}
\end{figure}

\begin{figure}
	\centering
	\includegraphics{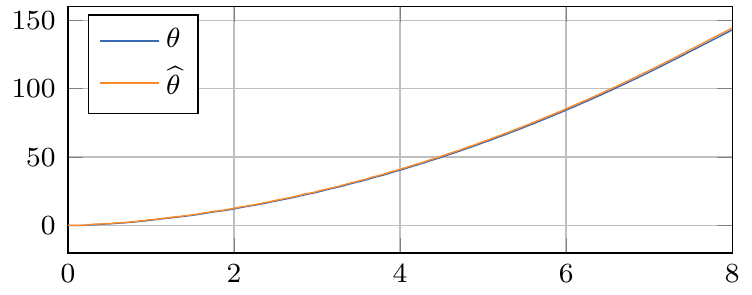}
	\includegraphics{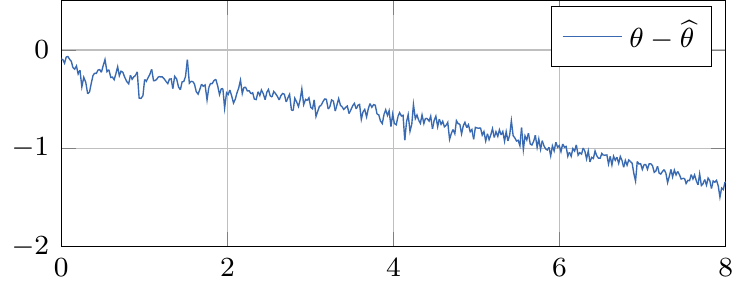}
	\caption{Reconstruction of $\theta$ in \si{\radian} (experiment).}
	\label{fig:xp_theta}
\end{figure}

The corresponding results on experimental data are shown in Fig.~\ref{fig:xp_cos_sin} and Fig.~\ref{fig:xp_theta}. Though of course not as good as in simulation, the agreement between the estimates and the ground truth is still very satisfying. The influence of magnetic saturation may account for part of the discrepancies. 

Fig.~\ref{fig:xp_zoom_s1albe} displays a close view of the ripple envelope~$s_1^{\alpha\beta}$ in approximately the same conditions as in Fig.~\ref{fig:s1ABsingleCarrier} when the rank of $\CA^{\alpha\beta}(u^{abc})$ is~2 case (top), and when the rank is~1 case with~$u_s^c=u_s^b\neq u_s^a$. They illustrate that though the experimental signals are distorted, they are nevertheless usable for demodulation.

\begin{figure}
	\centering
	\includegraphics{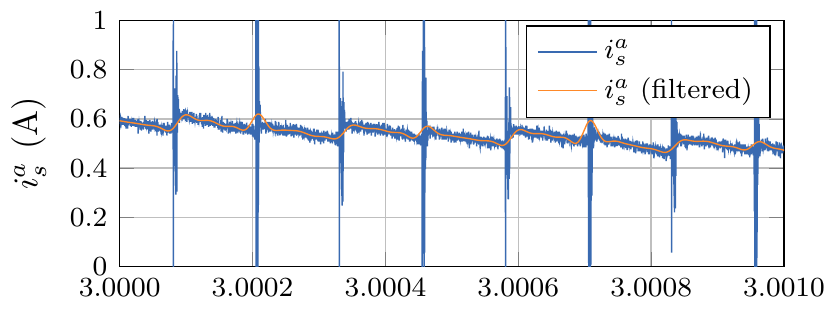}
	\caption{Measured current $i_s^a$ and its filtered version (experimental data).}
	\label{fig:current_filtered}
\end{figure}

Finally, we point out an important difference between the simulation and experimental data. In the experimental measurements, we notice periodic spikes in the current measurement, see figure~\ref{fig:current_filtered}; these are due to the discharges of the parasitic capacitors in the inverter transistors each time a PWM commutation occurs. As it might hinder the demodulation procedure of~\cite{CombeJMMR2016ACC,SurroopCMR2020arXiv}, the measured currents were first preprocessed
by a zero-phase (non-causal) moving average with a short window length of~$0.01\tpwm$. We are currently working on an improved demodulation procedure not requiring prefiltering.

\subsection{Interleaved PWM (simulation)}
\begin{figure}[t]
	\centering
	\hspace*{0.11cm}\includegraphics{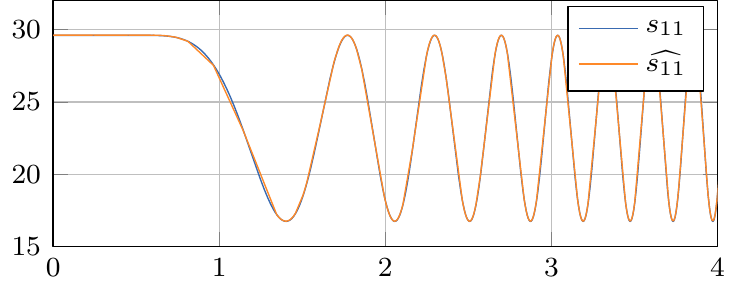}
	\includegraphics{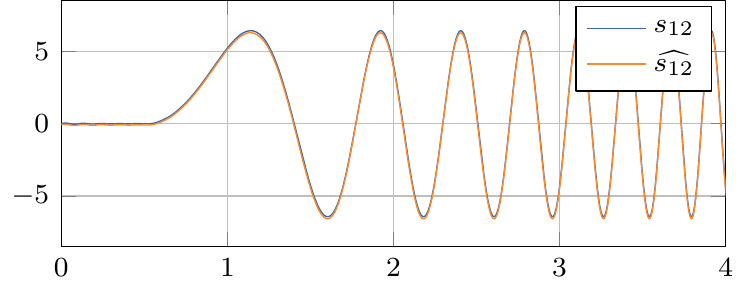}
	\includegraphics{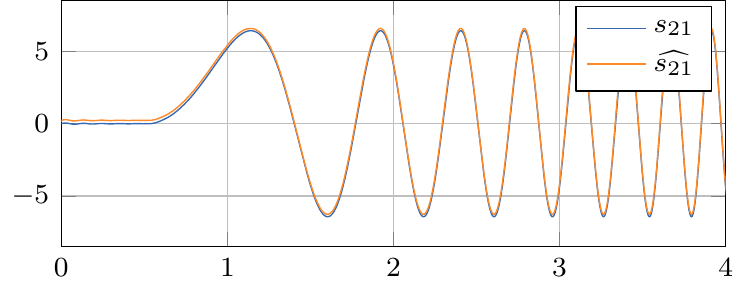}
	\includegraphics{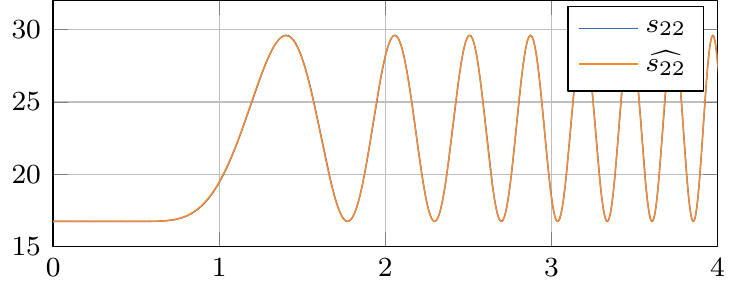}
	\caption{Reconstruction of saliency matrix $\CS(\theta)$ (simulation).}
	\label{fig:saliency_matrix}
\end{figure}

\begin{figure}
	\centering
	\includegraphics{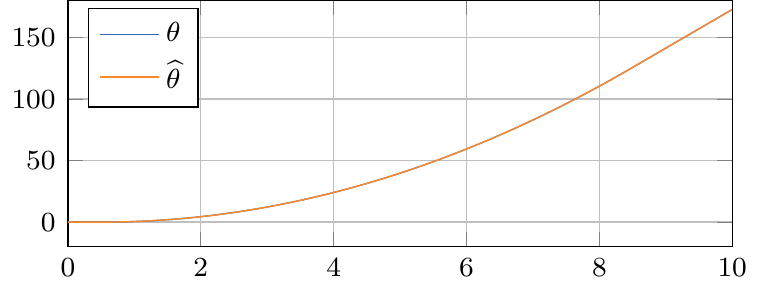}
	\includegraphics{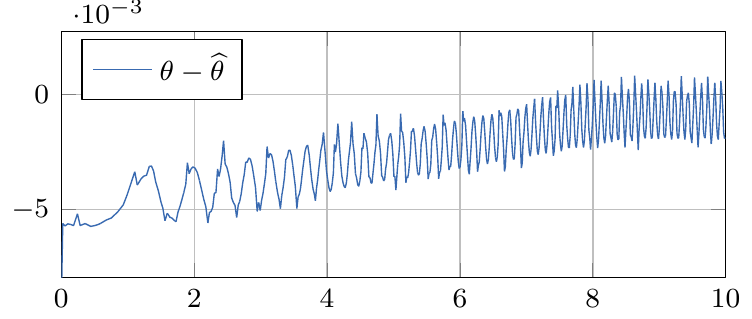}
	\caption{Rotor position $\theta$, its estimate $\widehat\theta$ (top); $\theta -\widehat\theta$ (bottom) (shifted carriers)}
	\label{fig:shifted_theta}
\end{figure}

The results obtained in simulation by the reconstruction procedure of section~\ref{subsec:shifted} for the saliency matrix and~$\CS(\theta)$ and for~$\theta$ are shown in Fig.~\ref{fig:saliency_matrix} and Fig.~\ref{fig:shifted_theta}. The agreement between the estimates and the actual values is excellent. We insit that the reconstruction does not require the knowledge of the magnetic parameters. 

\section{Conclusion}

This paper provides an analytic approach for the extraction of the rotor position of a PWM-fed PSMM, with signal injection provided by the PWM itself.  Experimental and simulations results illustrate the effectiveness of this technique. 

Further work includes a demodulation strategy not requiring prefiltering of the measured currents, and suitable for real-time processing. The ultimate goal is of course to be able to use the estimated rotor position inside a feedback loop.

\addtolength{\textheight}{-5.275cm}   


\bibliographystyle{phmIEEEtran}
\bibliography{biblio}
\end{document}